# Introducing Hierarchy in Energy Games

S. Lasaulce, Y. Hayel, R. El Azouzi, and M. Debbah

**Abstract**

In this work we introduce hierarchy in wireless networks that can be modeled by a decentralized multiple access channel and for which energy-efficiency is the main performance index. In these networks users are free to choose their power control strategy to selfishly maximize their energy-efficiency. Specifically, we introduce hierarchy in two different ways: 1. Assuming single-user decoding at the receiver, we investigate a Stackelberg formulation of the game where one user is the leader whereas the other users are assumed to be able to react to the leader's decisions; 2. Assuming neither leader nor followers among the users, we introduce hierarchy by assuming successive interference cancellation at the receiver. It is shown that introducing a certain degree of hierarchy in non-cooperative power control games not only improves the individual energy efficiency of all the users but can also be a way of insuring the existence of a non-saturated equilibrium and reaching a desired trade-off between the global network performance at the equilibrium and the requested amount of signaling. In this respect, the way of measuring the global performance of an energy-efficient network is shown to be a critical issue.

## I. Introduction

In this paper, we consider a decentralized multiple access channel (MAC). By definition [2], the MAC consists of a network of several transmitters and one receiver. The network is said to be decentralized in the sense that the receiver does not dictate to the users their transmit power level. Indeed, from the sole knowledge of his own uplink channel, each user can choose freely his power control policy in order to selfishly maximize a certain individual performance criterion, called utility (or payoff) in the context of game theoretic studies. The use of game theoretic tools is at the heart of the design of the recently advocated mobile flexible networks [3], which intend to break the spectral efficiency barrier through the use of intelligence. The selfish behavior enables to reduce the signaling overhead, especially for highly mobile terminals where topological information (channel state information -CSI- is one aspect) on the network can not be centralized. In this paper, unlike many works concerning this problem, the chosen users' utility is not the transmission rate (e.g., [4], [5], [6]) but the energy-efficiency of their communication. The latter approach, which consists in maximizing the ratio of the net number of information bits that are transmitted without error per time unit to the transmit power level, has been introduced in [7] for flat fading channels and recently re-used by [8] for multi-carrier CDMA (code division multiple access) systems and linear receivers, motivated by the facts that mobile terminals have a limited battery lifetime and in some applications (e.g., a sensor network measuring a temperature field) the main concern is not the transmission rate.

As mentioned in [7] the Nash equilibrium (NE) in such games can be energy inefficient. The NE of this power control game is shown to be Pareto inefficient. This is why [9] proposed, for MACs with flat fading links and single-user decoding (SUD), a pricing mechanism to obtain improvements in the users' utilities with respect to the case with no pricing. To our knowledge, since the release of [9], no alternative way of tackling this problem in the context of energy-efficient power control games has been proposed. In this paper we propose an alternative approach to [9] for improving the network efficiency by introducing a certain degree of hierarchy between the users. We propose two schemes. For the first scheme, we propose a Stackelberg formulation of the problem when SUD is assumed at the receiver. For the second scheme, we consider an a priori more efficient (and non-linear) receiver namely successive interference cancellation (SIC). Technically, our approach not only aims at improving the network equilibrium efficiency but has also two nice features: 1. It allows one to analyze the equilibrium uniqueness issue rigorously. Note that even in the simple case of linear pricing analyzed in [9], only simulations are provided to justify uniqueness; 2. Implementing pricing in real wireless networks is still an open issue for many well-used types of utilities (the problem being to know how to measure the modified utility) whereas energy-efficiency can be physically measured (for this purpose each terminal can evaluate its frame error rate over a certain period of time from a feedback mechanism and store the corresponding sequence of power levels); 3. Only individual CSI is needed at each transmitter in the regime of non-saturated equilibria, which is not case with pricing. More generally, our approach contributes to designing networks where intelligence is split between the base station (BS) and mobile stations (MSs) in order to find a desired trade-off between the global network performance reached at the equilibrium and the amount of signaling needed to make it work. As we will see, in both hierarchical approaches proposed the receiver only broadcasts common messages and the corresponding amount of additional signaling is reasonable. Note that the Stackelberg formulation arises naturally in some contexts of practical interest. For example, hierarchy is naturally present in contexts where there are primary (licensed) users and secondary (unlicensed) users who can sense their environment because there are equipped with a cognitive radio [12], [13], [14]. It is also natural if the users have access to the medium in an asynchronous manner. Note that there have been many works on Stackelberg games in the context of wireless communications [15], but they do not consider energy-efficiency for the individual utility as defined in [7], [8], [16]. Rather, they consider transmission rate-type utilities (see e.g., [5], [17], [18]).

This paper is structured as follows. The general signal model is provided in Sec. II-A. Sec. II-B reviews the main results of [8] for the non-cooperative game. Then, in Sec. III, we introduce our Stackelberg formulation by assuming an arbitrary choice for the game leader, when SUD is assumed. In Sec. IV we consider a different receiver namely a successive interference canceler, for which an arbitrary decoding order is chosen on each block of data (or packet). The choice of the best leader and decoding order in terms of overall network energy-efficiency is discussed in Sec. V. In Sec. VI we provide numerical results to illustrate the theoretical results derived in the previous sections. A short summary and some extensions of this work are provided in Sec. VII.

## II. Signal model and review of the non-cooperative game

### *A. Signal model*

Here we provide the signal model used in the whole paper except at the end of Sec. II-B where we explain how the provided results apply to the random CDMA case (RCDMA[1]). We consider a decentralized MAC with a finite number of users, which is denoted by $K$. We assume that the users transmit their data over block Rayleigh flat fading channels and the receiver knows on each block all the uplink channel gains (coherent communication assumption) whereas each transmitter has only access to the knowledge of its own channel. The latter assumption is realistic at least in two common scenarios: (a) the uplink-downlink channel reciprocity is valid and the BS sends training signals to the MSs; (b) the uplink channel varies slowly and the BS implements a reliable feedback mechanism to inform the MSs with their channel state. The equivalent baseband signal

[1]RCDMA: the entries of the spreading sequences correspond to the realizations of i.i.d random variables.

received by the base station can be written as

$$Y = \sum_{i=1}^{K} h_i X_i + Z \tag{1}$$

with $\forall i \in \{1, ..., K\}$, $\mathbb{E}|X_i|^2 = p_i$, $|h_i|$ is a Rayleigh distributed random variable and $Z \sim \mathbb{C}\mathcal{N}(0, \sigma^2)$. Each channel gain $h_i$ varies over time but is assumed to be constant over each block.

### B. Review of the non-cooperative game

In the system under investigation, users are selfish in the sense of their energy-efficiency. Here we review a few key results from [8], [9] concerning the non-cooperative game, which we will use to analytically evaluate the benefits brought by introducing hierarchy in this game. For any user $i \in \{1, ..., K\}$, the single-user signal-to-noise plus interference ratio (SINR) at the receiver writes as

$$\mathrm{SINR}_i = \frac{p_i |h_i|^2}{\sum_{j \neq i} p_j |h_j|^2 + \sigma^2} \tag{2}$$

where $j \neq i$. The strategy of user $i \in \{1, ..., K\}$ consists in choosing his transmit power level $p_i$ in order to maximize his utility function which is chosen to be:

$$u_i(p_1, ..., p_K) = \frac{T_i}{p_i} = \frac{R_i f(\mathrm{SINR}_i)}{p_i} \tag{3}$$

where $f$ is an efficiency function representing the packet success rate, which is assumed to be identical for all the users and $R_i$ is the transmission rate [7], [8] of user $i$. By definition of the utility (Eq. (3)) we see that the frequency at which the power control is updated is chosen to be the reciprocal of the data block duration. When it exists, the non-saturated NE[2] of this game is given by

$$\forall i \in \{1, ..., K\}, \ p_i^{\mathrm{SUD}} = \frac{\sigma^2}{|h_i|^2} \beta^* \mu^{\mathrm{SUD}} \tag{4}$$

where $\beta^*$ is the positive solution of the equation $xf'(x) = f(x)$ and $\mu^{\mathrm{SUD}} = \frac{1}{1-(K-1)\beta^*}$ is a penalty term due to multiple access interference; by using the term "non-saturated NE" we mean that the maximum transmit power for each user, denoted by $P_i^{\max}$, is assumed to be sufficiently high for not being reached at the equilibrium i.e., each user maximizes his energy-efficiency for a value less than $P_i^{\max}$. Several technical comments are in order. First, note that the equation $xf'(x) = f(x)$ has a positive solution if the function $f$ is sigmoidal [19] and verifies $u_i(0, \underline{p}_{-i}) = 0$, which is what assumed in this paper. Second, it has been shown [7] that there is always a unique NE in the game considered. If $\mu^{\mathrm{SUD}} \geq 0$ and none of the users' power constraints is saturated, the equilibrium is that given by Eq. (4). If one of the mentioned conditions is not met, that is if $\mu^{\mathrm{SUD}} < 0$ or at least one power constraint is saturated, the NE has to be rewritten by taking into account that some users transmit at their maximum power. What is important here is not to explicit the equilibrium in this case but to mention that the corresponding regime is less interesting for several reasons. We will only mention the most critical of them, which is not mentioned in the related works available in the literature ([7], [9], [8], etc.): for having such a saturated equilibrium the users need to know more than their own channel. For example in the $2-$user case, if user 2 transmits at his maximum power $p_2^{\mathrm{SUD}} = P_2^{\max}$ user 1 transmits at $p_1^{\mathrm{SUD}} = \beta^* \frac{\sigma^2}{|h_1|^2} + \beta^* \left|\frac{h_2}{h_1}\right|^2 P_2^{\max}$. Thus the interest in designing a system such that a non-saturated equilibrium is obtained is obvious. This is one of the reasons why we will assume such an operating regime in the whole paper. In this regime, even if a user has an infinite transmit power he will not necessarily use all of it. This is what Eq. (4) shows: each player tunes his transmit power in order for his SINR to be equal to $\beta^*$. In the sequel, we will denote by $u_i^{\mathrm{SUD}}$ the energy efficiency obtained by player $i$ at the NE.

[2]NE: the vector of strategies $\underline{p}^{\mathrm{SUD}} = (p_1^{\mathrm{SUD}}, ..., p_K^{\mathrm{SUD}})$ is an NE if $\forall i \in \{1, ..., K\}, \forall p_i \in [0, P_i^{\max}]$, $u_i(\underline{p}^{\mathrm{SUD}}) \geq u_i(p_i, \underline{p}_{-i}^{\mathrm{SUD}})$, with the standard notation $\underline{p}_{-i} = (p_1, ..., p_{i-1}, p_{i+1}, ..., p_K)$.

Note that the problem formulation presented in this paper can be applied to other types of systems, so our analysis is not exclusively applicable to the signal model defined by Eq. (1). For example, in flat fading RCDMA systems [20], the SINR after despreading (denoted by $\widetilde{\mathrm{SINR}}_i$) of the received signal can be written in the same form as Eq. (2):

$$\widetilde{\mathrm{SINR}}_i = \frac{\tilde{p}_i |h_i|^2}{\frac{\sum_{j \neq i} \tilde{p}_j |h_j|^2}{N} + \sigma^2} \tag{5}$$

where $N$ is the spreading factor (also the processing gain) of the RCDMA system. The strategy of user $i$ consists in choosing his transmit power level $\tilde{p}_i$ in order to maximize his utility function which is chosen to be:

$$\tilde{u}_i(\tilde{p}_1, \dots, \tilde{p}_K) = \frac{T_i}{\tilde{p}_i} = \frac{\tilde{R}_i \tilde{f}(\widetilde{\mathrm{SINR}}_i)}{\tilde{p}_i}. \tag{6}$$

The study of the case of RCDMA systems can be directly obtained from the signal model used in this paper by observing that the two models are merely linked by the following change of variables: $\tilde{t} = Nt$ where $t \in \{p_i, R_i, \mathrm{SINR}_i\}$ and $f(x) = \tilde{f}(Nx)$. This leads to $\tilde{u}_i(p_1, \dots, p_K) = \frac{\tilde{R}_i \tilde{f}(N.\mathrm{SINR}_i)}{\tilde{p}_i} = \frac{R_i f(\mathrm{SINR}_i)}{p_i}$. This clearly establishes the link between our signal model and that used by [8] for RCDMA systems and flat fading channels. Similarly it could also be linked to the case of RCDMA systems with frequency selective channels as recently shown in [21]. To conclude on RCDMA systems, it could be verified [8], [21] that the denominator of the transmit power at the NE becomes proportional to $1 - \frac{(K-1)\beta^*}{N}$, which is in favor of the existence of a non-saturated equilibrium in games under investigation.

### C. Information assumptions

To have a clear view of what is assumed to be known at which terminal, we mention here all our assumptions in terms of information for all the terminals. As the communications are assumed to be coherent, the BS knows in all the cases treated in this paper all the channel gains $h_1, ..., h_K$ on each data block. As it will be seen, for the two hierarchical games introduced in this work, each user (say user $i$) only needs to know his own channel (i.e., $h_i$) to implement his optimal selfish power control policy. We are therefore in the same situation as the non-cooperative game. On the other hand if full CSI $h_1, ..., h_K$ would be available at each transmitter, it would be possible to formulate the power control problem as a cooperative game (team game) with a common utility, which is not considered in this paper. To illustrate this point, in the $2-$user case it can be checked that the best set of SINRs for the team, denoted here by $(x_i)_i$, is the solution of the system of equations $\forall i \in \{1,2\}, x_i f'(x_i) - f(x_i) - f'(x_{-i}) \left[ \left| \frac{h_{-i}}{h_i} \right| \frac{1+x_{-i}}{1+x_i} \right]^2 x_i^2 = 0$. We will also assume a context of games with complete information that is, each user perfectly knows the game (the number of users, the sets of strategies and the different utilities) and every user is assumed to be rational [22]. The additional assumption we make w.r.t. [7], [8] is that: in the Stackelberg formulation with SUD, there exists a mechanism that allows the followers to know the (receive) power level of the leader, which can be acquired with an appropriate sensing system or by assuming that the BS sends an appropriate broadcast signal; in the game with SIC all the users know the decoding order used by the BS.

## III. A HIERARCHICAL GAME WITH SINGLE USER DECODING

As mentioned previously, one of our motivations for introducing hierarchy is to improve the network energy-efficiency. The proposed approaches can be seen as intermediate schemes between the totally centralized power control policy and the non-cooperative policy of [7], [8]. It is also quite relevant for flexible networks where the trend is to split the intelligence between the network infrastructure and the (generally mobile) users' equipments. These approaches are therefore reasonable ways of finding a desired trade-off between the desired global network performance and the amount of control signaling sent by the receiver. In this section, we propose a Stackelberg formulation of the power control game where one of the $K$ users is chosen to be the leader whereas the others are the followers. The receiver is not a player of the game here. In this respect we will always assume in this section that SUD is implemented at the receiver. The motivations for using SUD can be precisely that the receiver has to remain neutral in the game, or/and for limiting the receiver complexity, or/and to minimize the possible signaling cost induced by a more advanced receiver. In Sec. V however, we will use a more advanced receiver than SUD namely SIC, which naturally introduces hierarchy between users via the decoding order used by the BS. In the case where this order is optimized, the problem can be formulated as a Stackelberg game where the receiver has his own utility (similarly to [5] where Shannon transmission rates are considered for the users' utilities) and is the game leader.

Here, we consider without loss of generality (but possibly with loss of optimality) that user $i$ is the leader of the game. Even though there is no loss of generality mathematically speaking, this arbitrary choice might seem to be artificial physically. In fact, there are useful scenarios where some terminals are naturally leaders of the game, by definition of the context. For instance, in wireless networks with primary and secondary users, most often only secondary terminals are equipped with a cognitive radio and can be followers; the primary terminals which have been generally deployed in the first place are therefore leaders of the game by conception of the network. Back to our system model, each follower $j \neq i$ plays a non-cooperative game with the other followers, given what the leader plays. Interestingly, it is possible to show that, under realistic conditions, there is a unique equilibrium in this hierarchical game, which is called a Stackelberg Equilibrium (SE). Before indicating how the users determine their optimal transmit power, let us define a Stackelberg equilibrium. Let $\mathcal{U}^*(p_i)$ be the set of NE for the group of followers when the leader plays strategy $p_i$. In other words, the leader maximizes his utility function which depends on the NE $u^* \in \mathcal{U}^*(p_i)$ of the followers.

*Definition 1 (Stackelberg equilibrium): A vector* $\underline{p}^{\mathrm{SE}} = (p_i^{\mathrm{SE}}, p_{-i}^{\mathrm{SE}})$ *is called a Stackelberg equilibrium (SE) if* $\underline{p}_{-i}^{\mathrm{SE}} \in \mathcal{U}^*(p_i)$ *and the power* $p_i^{\mathrm{SE}}$ *maximizes the utility function of user* $i$*, the leader of the game. In other words* $p_i^{\mathrm{SE}} = \arg\max_{\tilde{p}_i} u_i(\tilde{p}_i, \underline{p}_{-i}^{\mathrm{SE}}(\tilde{p}_i))$.

By denoting $(p_i^{\mathrm{SE}}, \underline{p}_{-i}^{\mathrm{SE}})$ the power profile at the SE, this definition translates mathematically by

$$p_i^{\mathrm{SE}} = \arg\max_{p_i} u_i\left(p_1^{\mathrm{SE}}(p_i), \ldots, p_{i-1}^{\mathrm{SE}}(p_i), p_i, p_{i+1}^{\mathrm{SE}}(p_i), \ldots, p_K^{\mathrm{SE}}(p_i)\right), \tag{7}$$

where $p_j^{\mathrm{SE}}(p_i)$, $j \neq i$, is the power of the follower $j$ at the NE. Now we look at the problem of the existence and uniqueness of such a vector of transmit power levels. A solution to these issues is stated through the following proposition.

*Proposition 2:* (EXISTENCE AND UNIQUENESS OF AN SE). *There is a unique Stackelberg equilibrium* $\underline{p}^{\mathrm{SE}}$ *in the proposed hierarchical game where user* $i$ *is the leader:*

$$p_i^{\mathrm{SE}} = \frac{\sigma^2}{|h_i|^2} \frac{\gamma^*(1+\beta^*)}{1-(K-1)\gamma^*\beta^* - (K-2)\beta^*} \tag{8}$$

*and for each follower* $j \neq i$

$$p_j^{\mathrm{SE}} = \frac{\sigma^2}{|h_j|^2} \frac{\beta^*(1+\gamma^*)}{1-(K-1)\gamma^*\beta^* - (K-2)\beta^*}, \tag{9}$$

*if the following (sufficient) conditions hold:* $\frac{f''(0)}{f'(0)} \geq 2\frac{(K-1)\beta^*}{1-(K-2)\beta^*}$ *and* $\phi(x) = x\left[1 - \frac{(K-1)\beta^*}{1-(K-2)\beta^*}x\right]f'(x) - f(x)$ *has a single stationary point in* $]0, \gamma^*[$, *where where* $\beta^*$ *is the positive solution of the equation* $xf'(x) - f(x) = 0$ *and* $\gamma^*$ *is the positive solution of the equation* $\phi(x) = 0$.

This proposition is proven in Appendix A. In order to have an idea to what extent the sufficient conditions stated in Prop. 2 are realistic, we consider the practical choice of efficiency function proposed by [7] and also used by [8][16]: $f(x) = (1 - e^{-x})^M$ where $M$ is the block length (this function is a reasonable model for evaluating the packet success rate of a transmission). Assuming this function, we have $\phi'(x) = e^{-x}\left\{\frac{(K-1)\beta^*}{1-(K-2)\beta^*}x^2 - \left[2\frac{(K-1)\beta^*}{1-(K-2)\beta^*}M + 1\right]x + M - 1\right\}$ and the existence and uniqueness of an SE readily follows (see Appendix A). Commenting the result of Prop. 2 itself, an interesting feature of the SE can be noticed. The SE has the same attractive property as the equilibrium of the non-cooperative game of [7] namely each user only needs to know his own channel to do what is best for him. This result comes from the facts thats that all users are rational, every user knows that the others are rational and more specifically every user knows how the others are going to tune their SINR, which is identical for all the followers. At this point, some important questions arise. From a user point of view, is it better to be chosen to be a leader or a follower? With respect to the non-cooperative game what is the gain brought by introducing hierarchy? Do all the players benefit from this? The first question is answered in Prop. 3 whereas the latter questions are the purpose of Prop. 4.

*Proposition 3:* (FOLLOWING IS BETTER THAN LEADING) *Every user has always a better utility by being chosen as a follower instead of a leader*.

*Proof:* We denote by $u_L^{\text{SE}}$ (resp. $u_F^{\text{SE}}$) the utility of user $i \in \{1\ldots, K\}$ (resp. $j \neq i$) when he is chosen to be the leader (resp. a follower) of the game. First, we observe that, at the Stackelberg equilibrium, the SINR for the leader and a follower are: $\text{SINR}_L^{\text{SE}} = \gamma^*$ and $\text{SINR}_F^{\text{SE}} = \beta^*$. From [19], we have that for all $x > 0$: $x > \beta^* \Leftrightarrow xf'(x) < f(x)$. As for all $x > 0$, $x\left[1 - \frac{(K-1)\beta^*}{1-(K-2)\beta^*}x\right]f'(x) < xf'(x)$, from a simple geometrical argument we see that $\gamma^* < \beta^*$ (the fact that the leader gets a lower SINR than in the non-cooperative game, and therefore lower than the follower, can also be understood by noticing that $\text{SINR}_1$ is a function of $p_1$ only and grows more slowly with $p_1$). This means that the SINR of the follower (i.e., $\beta^*$) is higher than the SINR of the leader (i.e., $\gamma^*$). Therefore, we can write that

$$\frac{u_L^{\text{SE}}}{u_F^{\text{SE}}} = f(\gamma^*)\frac{|h_i|^2}{\sigma^2}\frac{1-(K-1)\beta^*\gamma^* - (K-2)\beta^*}{\gamma^*(1+\beta^*)}\frac{1}{f(\beta^*)}\frac{\sigma^2}{|h_i|^2}\frac{\beta^*(1+\gamma^*)}{1-\beta^*(K-1)\gamma^* - (K-2)\beta^*} \quad (10)$$

$$= \frac{\frac{f(\gamma^*)}{\gamma^*}}{\frac{f(\beta^*)}{\beta^*}}\frac{1+\gamma^*}{1+\beta^*} = \frac{g(\gamma^*)}{g(\beta^*)}\frac{1+\gamma^*}{1+\beta^*} \leq 1, \quad (11)$$

where the inequality follows from $\gamma^* \leq \beta^*$ and the fact that function $g : x \mapsto \frac{f(x)}{x}$ reaches its maximum in $\beta^*$. ■

The main issue we need to address now is the comparison between the non-cooperative and hierarchical games in terms of energy efficiency. Specifically, we want to compare the values of $u_i(\underline{p}^{\mathrm{SE}})$ and $u_i(\underline{p}^{\mathrm{SUD}})$, for each player $i \in \{1, \ldots, K\}$. This is stated through the following proposition.

*Proposition 4:* (UNIFORM IMPROVEMENT OF UTILITIES). *We always assume that SUD is used at the BS. Then, both the leader and followers improve their utility with respect to the non-cooperative setting.*

*Proof:*

(a) All the followers improve their utility. By denoting $j$ the index of a given follower we have:

$$\frac{u_j^{\mathrm{SE}}}{u_j^{\mathrm{SUD}}} = f(\beta^*)\frac{|h_j|^2}{\sigma^2}\frac{1-(K-1)\beta^*\gamma^*-(K-2)\beta^*}{\beta^*(1+\gamma^*)}\frac{1}{f(\beta^*)}\frac{\sigma^2}{|h_j|^2}\frac{\beta^*}{1-(K-1)\beta^*} \tag{12}$$

$$= \frac{1-(K-1)\beta^*+\beta^*\left[1-(K-1)\gamma^*\right]}{(1+\gamma^*)\left[1-(K-1)\beta^*\right]} = \frac{1+\beta^*-(K-1)\beta^*(1+\gamma^*)}{1+\gamma^*-(K-1)\beta^*(1+\gamma^*)}. \tag{13}$$

We see that this ratio is higher than 1 since $\gamma^* \leq \beta^*$.

(b) The leader improves his utility. Denoting by $i \neq j$ the index of the leader we have:

$$\frac{u_i^{\mathrm{SE}}}{u_i^{\mathrm{SUD}}} = f(\gamma^*)\frac{|h_i|^2}{\sigma^2}\frac{1-(K-1)\beta^*\gamma^*-(K-2)\beta^*}{\gamma^*(1+\beta^*)}\frac{1}{f(\beta^*)}\frac{\sigma^2}{|h_i|^2}\frac{\beta^*}{1-(K-1)\beta^*} \tag{14}$$

$$= \frac{\frac{f(\gamma^*)}{\gamma^*}}{\frac{f(\beta^*)}{\beta^*}}\frac{1-\beta^*\left[K-2+(K-1)\gamma^*\right]}{1-\beta^*\left[K-2+(K-1)\beta^*\right]} = \frac{h(\gamma^*)}{h(\beta^*)}, \tag{15}$$

where $h(x) = \frac{f(x)}{x}\left[1-\beta^*(K-2+(K-1)x)\right]$. It can be checked that

$$h'(x) = \frac{x\left[1-\frac{(K-1)\beta^*}{1-(K-2)\beta^*}x\right]f'(x)-f(x)}{x^2} = \frac{\phi(x)}{x^2}, \tag{16}$$

where $\phi$ is defined in Prop. 2. We have $\phi(\gamma^*) = 0$, $\phi(\beta^*) < 0$ and as the solution of $\phi(x) = 0$ is unique, for all $x \in [\gamma^*, \beta^*]$ we have $\phi(x) < 0$. The function $h$ is therefore non-increasing over $[\gamma^*, \beta^*]$ and we have $u_i^{\mathrm{SE}} \geq u_i^{\mathrm{SUD}}$ when user $i$ is the leader, which concludes the proof. ■

To conclude this section we will make two comments. First, it is very interesting to observe that all the players benefit from hierarchy when energy-efficiency is chosen to be the users' utility. This result is not usual in game-theoretic studies. In economics, for instance, in the case of a duopoly [23], only the leader can benefit from the introduction of hierarchy. Even more surprisingly, in our context, a user prefers to be a follower than a leader. In cognitive networks with primary and secondary users, only terminals that are equipped with a cognitive radio will have this privilege. In a conventional cellular network, the designation of a leader can be made by the By broadcasting $p_1$ the base station discloses the user who allows a certain global performance metric to be maximized (this will be the purpose of Sec. V) and the power level he should play. As the users, who are designated as followers, are selfish, rational, and can not coordinate each other they are going to play their best response $p_i(p_1)$. Knowing this, the user who is designated as a leader has to transmit at $p_1$ to maximize his utility. As a second comment, we note that all the users not only obtain a better energy-efficiency in the proposed Stackelberg game but it can also be checked that they transmit with a lower power than in the non-cooperative game (Sec. II-B), which is in favor of creating a network where interference is self-regulated by the users.

## IV. A Hierarchical Game with Successive Interference Cancellation

The approach presented in this section is clearly related to that of Sec. III in the sense that it also consists in using hierarchy to improve the network equilibrium efficiency. This approach, based on the use of SIC at the BS is even more strongly related to Sec. III in the case where the BS has his own utility, since we have a Stackelberg in this case. On the other hand, this section and Sec. III corresponds to two different points of view: Sec. III corresponds more to a game theoretic standpoint for which the receiver is unchanged but hierarchy is introduced between players (and therefore influencing their transmission strategy) to make the society more efficient while the second approach (using SIC) typically corresponds to what a wireless engineer could do in order to improve the network performance that is to say implement a more advanced receiver (SUD $\rightarrow$ SIC). From now on, let us assume that the BS implements successive interference cancellation. The principle of SIC is to rank the users and decode them successively (see e.g., [2]). For the 2-user case the decoding is a two-stage process. In the first stage, the receiver decodes a user (say user 1) by considering the other (user 2) as part of the noise. In the second stage, i.e., after the first user has been decoded successfully, the first user can be subtracted from the received signal and user 2 is decoded without multiuser interference. Compared to the case of the SUD-based receiver, the SIC-based receiver does not require any additional knowledge and therefore always only uses the channel state information $(h_1, ..., h_K)$ on each block of data, just as SUD. From a practical point of view the two main differences between SIC and SUD is that SIC is more complex to be implemented and the decoding order has to be known to the users. For the latter point, as mentioned in [6], it does not necessarily mean that the receiver has to send a signal for indicating the decoding order to the user. In fact, this information can also be acquired from an external (and therefore free in terms of signaling cost) source of signal. However, in this case there is generally a loss of optimality for the overall network performance. Clearly, one of the main advantages for using a SIC at the receiver is to partially remove some multiuser interference. Note that, in the context of systems with mutual interaction among users, improving the decoding scheme does not necessarily imply that each user improves his utility (Braess-like paradoxes [24] can sometimes occur) . All the issues we have just mentioned are precisely the purpose of this section.

*Proposition 5:* (EXISTENCE AND UNIQUENESS OF AN NE). *Let denote by $i$ the index of the user who is decoded with rank $K-i+1$ in the successive decoding procedure at the receiver. In the non-cooperative game with a SIC-based receiver where the utility is chosen to be given by Eq. (3) where the SINRs are those considered at the* ***output*** *of the SIC, there exists a unique (pure) NE $(p_1^{\mathrm{SIC}}, ..., p_K^{\mathrm{SIC}})$ which is given by:*

$$\forall i \in \{1, ..., K\},\ p_i^{\mathrm{SIC}} = \frac{\sigma^2}{|h_i|^2} \beta^* \mu_i^{\mathrm{SIC}} \tag{17}$$

where $\mu_i^{\mathrm{SIC}} = (1+\beta^*)^{i-1}$ *is a penalty term due to multiple access interference.*

*Proof:* The existence of an NE is insured by the geometrical and topological properties of the utility functions and strategy sets of the users, over which the maximization is performed. Indeed, since for every user $i$, the utility $u_i$ is continuous in $\underline{p} = (p_1, ..., p_K)$ and is quasi-concave w.r.t. to $p_i$ over the convex and compact strategy sets $[0, P_1^{\max}]$, ..., $[0, P_K^{\max}]$, we can apply Debreu-Fan-Glicksberg theorem (see e.g., [25]). This guarantees the existence of at least one pure NE. To prove the uniqueness of the NE we apply a result derived by [26] and more recently re-used by [27]. We know from [26] that if the best response (BR) correspondence $\underline{\mathrm{BR}}(\underline{p}) = \left(\mathrm{BR}_1(\underline{p}), ..., \mathrm{BR}_K(\underline{p})\right)$ is monotonic and scalable, then the NE is unique. Although we deal here with a non-linear receiver we see that the SINR of user $i$ is given by

$$\mathrm{SINR}_i = \beta^* = \frac{p_i |h_i|^2}{\sigma^2 + \sum_{j=0}^{i-1} |h_j|^2 p_j}. \tag{18}$$

(with the notational convention $p_0 = 0$) and has the same key property as the SINR obtained with linear receivers [8][16] i.e., $p_i \frac{\partial \mathrm{SINR}_i}{\partial p_i} = \mathrm{SINR}_i$. Thus, in order to maximize his utility (Eq. (3)) each user has to tune his transmit power such that his SINR equals to $\beta^{\mathrm{SIC}} = \beta^*$ where $\beta^* f'(\beta^*) = f(\beta^*)$. Knowing this, it is easy to express the best responses of the users. Here we have that $\forall \in \{1, ..., K\},\ \mathrm{BR}_i(\underline{p}) = \frac{\beta^*}{|h_i|^2}\left(\sigma^2 + \sum_{j=0}^{i-1} p_j |h_j|^2\right)$. According to our assumptions, these expressions of the BRs are valid in the non-saturated regime, otherwise they can be equal to $P_i^{\max}$ for user $i$. Clearly we have that: 1. $\underline{p} \geq \underline{p}' \Rightarrow \underline{\mathrm{BR}}(\underline{p}) \geq \underline{\mathrm{BR}}(\underline{p}')$ since $\frac{\partial \mathrm{BR}_i(\underline{p})}{\partial p_j} = \beta^* \left|\frac{h_j}{h_i}\right|^2 \geq 0$ (monotonicity); 2. $\forall \alpha > 1,\ \alpha \underline{\mathrm{BR}}(\underline{p}) > \underline{\mathrm{BR}}(\alpha \underline{p})$ (scalability). At last, in our context of games with complete information, the expressions of the transmit powers at the equilibrium directly follow from the aforementioned property and expressions of the SINRs. ■

At least two key points are worth being noticed here. First, we see that, in contrast with the NE and SE with SUD, the existence of a non-saturated NE is still insured when $\beta^* > \frac{1}{K-1}$. Second, it is important to note that Prop. 5 indicates that, at the equilibrium, the transmit power of a user grows exponentially with $i$, which is related to his decoding rank (say $d_i$) by $d_i = K - i + 1$. As the penalty term for SUD (see Eq. (4)) is an hyperbolic function of $(K-1)\beta^*$, this seems to indicate that SIC might be less energy-efficient than SUD. It turns out that this is not the case in the regime where non-saturated equilibria exist for the non-cooperative game. This is the purpose of the following proposition.

*Proposition 6:* (SIC VERSUS SUD).*Let always denote by $i$ the index of the user who is decoded with rank $K-i+1$ in the successive decoding procedure at the receiver. Then, every user prefers to be in the game with a receiver implementing SIC instead of the game with a SUD-based receiver.*

*Proof:* First, let us prove that the sequence defined by $\rho_i = \frac{u_i^{\mathrm{SIC}}}{u_i^{\mathrm{SUD}}}$ is non-increasing. The ratio of the utility of user $i$ at the NE when SUD is assumed (non-cooperative game of Sec. II-B) to that obtained when SIC is assumed is:

$$\begin{aligned} \rho_i &= \frac{R_i f(\beta^*)}{p_i^{\mathrm{SIC}}} \frac{p_i^{\mathrm{SUD}}}{R_i f(\beta^*)} && (19) \\ &= \frac{p_i^{\mathrm{SUD}}}{p_i^{\mathrm{SIC}}} && (20) \\ &= \frac{1}{[1-(K-1)\beta^*](1+\beta^*)^{i-1}}. && (21) \end{aligned}$$

Clearly we have that $\forall i \in \{1, ..., K-1\}$, $\frac{\rho_{i+1}}{\rho_i} = 1 + \beta^* \geq 1$, which, in particular, mathematically translates the fact that the user who is the less likely to prefer SIC is user $K$ since he is decoded first.

Now let us prove that user $K$ has a better utility with SIC than with SUD. For this purpose we need to prove that $\rho_K \geq 1$. By defining the function $\rho_K : x \mapsto \frac{1}{[1-(K-1)x](1+x)^{K-1}}$ we have that

$$\frac{\partial \rho_K}{\partial x} = \frac{K(K-1)x(1+x)^{K-2}}{\{[1-(K-1)x](1+x)^{K-1}\}^2}. \qquad (22)$$

As this derivative is non-negative in the interval of non-saturated equilibria for the non-cooperative game i.e., for $x \in \left[0; \frac{1}{K-1}\right[$ and $\rho_K(0) = 1$, we therefore have that $\rho_K \geq 1$ in the interval of interest. ■

The proven result translates the fact that in the game with SIC, users are more energy-efficient and more specifically, transmit with a lower power. Therefore, every user sees less interference than in the game with SUD. In particular, user $K$ sees $K-1$ interferers in both games but the amount of interference he undergoes is less when SIC is implemented.

## V. NETWORK ENERGY-EFFICIENCY ANALYSIS

So far, in the two hierarchical games analyzed, we have assumed an arbitrary choice for the follower (Stackelberg game with SUD) and decoding order (non-cooperative game with SIC). In this section we want to assess the influence of these degrees of freedom on the overall network energy-efficiency. For this purpose, we consider two measures: the social welfare [28] which is well known in game theoretic studies and the energy-efficiency of the equivalent virtual multiple input multiple output (MIMO) system, the latter being used as individual utility to optimize power allocation in multicarrier CDMA systems [8]. This will allow us to have two complementary points of view on the way of measuring energy-efficiency for a network. As a comment regarding the terminology used, note that if the non-cooperative game with SIC is optimized in terms of a certain measure of energy-efficiency of the global network, the game can also be seen as a Stackelberg game where: (a) the receiver is the game leader; (b) the users (mobile terminals) are the followers; (c) his set of strategies is the set of all decoding orders; (d) his utility is $w$ (Eq. (23)) or $v$ (Eq. (30)). In the case of the Stackelberg game with SUD, if the BS has his own utility, we will referred to it as the super-leader to avoid confusion with the MS that is already called leader.

### *A. Social welfare*

The social welfare of the network is measured by the total utility of the system, which is expressed as follows:

$$w = \sum_{i=1}^{K} u_i = \sum_{i=1}^{K} \frac{T_i}{p_i}. \tag{23}$$

For this measure we have the following two results.

*Proposition 7:* (BEST CHOICE OF THE LEADER). *Assume a Stackelberg game with SUD. In order to maximize the social welfare, the user who has the lowest $R_i|h_i|^2$ has to be chosen as the game leader.*

*Proof:* Let $w^{(i)}$ be the social welfare when user $i$ is chosen to be the leader (resp. follower) and $p_i^L$ (resp. $p_i^F$) be his transmit power at the SE. We have that

$$\begin{aligned} w^{(i)} - w^{(j)} &= R_i \frac{f(\gamma^*)}{p_i^L} + R_j \frac{f(\beta^*)}{p_j^F} - R_i \frac{f(\beta^*)}{p_i^F} - R_j \frac{f(\gamma^*)}{p_j^L} && (24) \\ &= \frac{(R_j|h_j|^2 - R_i|h_i|^2)}{|h_i|^2} \left[ \frac{f(\beta^*)}{p_i^F} - \frac{f(\gamma^*)}{p_i^L} \right] && (25) \\ &= \frac{(R_j|h_j|^2 - R_i|h_i|^2)}{|h_i|^2} \left[ u_{i,F}^{\mathrm{SE}} - u_{i,L}^{\mathrm{SE}} \right] > 0 && (26) \end{aligned}$$

where equality (25) follows from $|h_i|^2 p_i^L = |h_j|^2 p_j^L$ and $|h_i|^2 p_i^F = |h_j|^2 p_j^F$. From Prop. 3, any user has always a better utility by being chosen as a follower instead of a leader, then we see that the difference is non-negative if and only if $R_i|h_i|^2 \leq R_j|h_j|^2$, which concludes the proof. ■

*Proposition 8:* (BEST DECODING ORDER). *Assume a non-cooperative game with SIC. The best decoding order in the sense of the social welfare is to decode the users in the increasing order of their energy weighted by the coding rate* $R_i|h_i|^2$.

*Proof:* Let $\pi \in \mathcal{P}$ be the permutation operator corresponding to the choice of the decoding order. Since the users have the same SINR at the equilibrium we have that:

$$\begin{aligned}\pi_w &= \arg\max_{\pi\in\mathcal{P}} w^{(\pi)} &(27)\\ &= \arg\max_{\pi\in\mathcal{P}} f(\beta^*)\sum_{i=1}^{K}\frac{R_i}{p_i} &(28)\\ &= \arg\max_{\pi\in\mathcal{P}} \sum_{i=1}^{K} R_i|h_i|^2 \times \frac{1}{(1+\beta^*)^{i-1}}. &(29)\end{aligned}$$

As by definition $\beta^* > 0$, the desired result follows. ■

To have more insights on the two derived results let us consider the 2-user case. From these two propositions we see that using the social welfare as a global measure of efficiency always gives an advantage to the dominant user in asymmetric channels i.e., for which $|h_i|^2 >> |h_j|^2$ for $i \neq j$. Indeed, in the Stackelberg game with SUD the strongest user is chosen to be the follower and in the non-cooperative game with SIC he is chosen to be decoded last. In the limit case where $\frac{|h_1|}{|h_2|} \to +\infty$, we have that $\lim_{\frac{|h_1|}{|h_2|}\to+\infty} w = +\infty$. Therefore if one user becomes more and more satisfied the whole society becomes more and more satisfied. Clearly, the social welfare as defined by Eq. (23) is not that social one could expect in the sense that it ignores fairness. This is one of the reasons why other measures of efficiency have to be used in certain scenarios. In the next subsection we propose to consider the efficiency of the equivalent virtual MIMO system.

### *B. Equivalent virtual MIMO network (EVMN) energy-efficiency*

We consider now another performance metric which corresponds to the energy-efficiency of an equivalent system where the transmitters would be co-located. It is described in [8] as the individual utility in the context of multi-carrier systems:

$$v = \frac{\sum_{i=1}^{K} T_i}{\sum_{i=1}^{K} p_i}. \tag{30}$$

It turns out that one can obtain very different conclusions by optimizing this quantity with respect to the degrees of freedom instead of the social welfare.

*Proposition 9:* (BEST CHOICE OF THE LEADER). *Assume a Stackelberg game with SUD. Without loss of generality assume that* $|h_1|^2 \leq |h_2|^2 \leq ... \leq |h_K|^2$. *In order to maximize the EVMN energy-efficiency, user* $i$ *has to be chosen as the game leader if*

$$v^{(j)} \geq \frac{(R_i - R_j)\left[f(\beta^*) - f(\gamma^*)\right]\left[1 - (K-1)\beta^*\gamma^* - (K-2)\beta^*\right]}{\sigma^2(\beta^* - \gamma^*)\left(\frac{1}{|h_j|^2} - \frac{1}{|h_i|^2}\right)} \text{ for } j \leq i, \text{ and} \tag{31}$$

$$v^{(j)} \leq \frac{(R_i - R_j)\left[f(\beta^*) - f(\gamma^*)\right]\left[1 - (K-1)\beta^*\gamma^* - (K-2)\beta^*\right]}{\sigma^2(\beta^* - \gamma^*)\left(\frac{1}{|h_j|^2} - \frac{1}{|h_i|^2}\right)} \text{ for } j \geq i. \tag{32}$$

The proof of this result is provided in Appendix B. To illustrate this proposition, let us consider the 2-user case. Without loss of generality assume that $|h_i|^2 < |h_j|^2,\ i \neq j$. In this case one can check that in order to maximize the EVMN user $i$ (resp. $j$) has to be chosen as the game leader if $aR_i \leq R_j$ (resp. $aR_i \geq R_j$) where $a < 1$ is defined by $a \triangleq \frac{f(\beta^*)\alpha_j - f(\gamma^*)\alpha_i}{f(\beta^*)\alpha_i - f(\gamma^*)\alpha_j}$ with $\alpha_i = |h_i|^2\gamma^*(1+\beta^*) + |h_j|^2\beta^*(1+\gamma^*)$ and $\alpha_j = |h_j|^2\gamma^*(1+\beta^*) + |h_i|^2\beta^*(1+\gamma^*)$.

Now let us turn our attention to the hierarchical game with SIC where the BS has to rank the different users to maximize the EVMN.

*Proposition 10:* (BEST DECODING ORDER). *Assume a non-cooperative game with SIC. The best decoding order in the sense of the EVMN energy-efficiency is to decode the users in the decreasing order of their signal-to-noise ratio (SNR)* $\frac{|h_i|^2}{\sigma^2}$.

*Proof:* Let $\pi \in \mathcal{P}$ be the permutation operator corresponding to the choice of the decoding order. Since the users have the same SINR at the equilibrium we have that $v = \frac{f(\beta^*) \times \sum_{i=1}^{K} R_i}{\sum_{i=1}^{K} p_i^{\mathrm{SIC}}}$. Therefore,

$$\pi_v = \arg\max_{\pi \in \mathcal{P}} v^{(\pi)} = \arg\min_{\pi \in \mathcal{P}} \sum_{i=1}^{K} p_i^{\mathrm{SIC}}. \tag{33}$$

As $p_i^{\mathrm{SIC}} = \frac{\sigma^2}{|h_i|^2}\beta^*(1+\beta^*)^{i-1}$ it is clear that the influence of a user on the sum of powers decreases with his decoding rank. To minimize the total system power one has to decode the users with a decreasing order of their SNR. ■

Here again, in order to have insights on the addressed issue, let us consider the 2-user case. From this proposition we see that maximizing $v$ over the choice of decoding order amounts to giving more to the poorest user in terms of $|h_i|^2$ whereas maximizing $w$ w.r.t the follower/leader choice amounts to giving more to the richest user in terms of $R_i|h_i|^2$. Also, in the case of asymmetric MACs the conclusions are markedly different from those obtained with $w$. We already know that $\lim_{\frac{|h_1|}{|h_2|} \to +\infty} w = +\infty$. An equivalent of $v$ when $\frac{|h_1|}{|h_2|} \to +\infty$ is $v \sim \frac{|h_2|^2}{\sigma^2}\beta^*(R_1+R_2)f(\beta^*)$. Now, with the latter measure of energy-efficiency, even if a user gets very rich, the wealth of the whole society does not increase and is limited by the poorest user.

## VI. Numerical examples

First, we want to analyze the performance of a network for which the receiver implements SIC. For this purpose we assume the following scenario: $K = 10$, $M = 100$, $N = 1$ (no spreading), $R_i = 100$ kbps for all $i \in \{1, ..., K\}$ and $E|h_i|^2 = 1$ for all $i \in \{1, ..., K\}$; the efficiency-function chosen is $f(x) = (1 - e^{-x})^M$. Fig. 1 and 2 respectively represent the network energy-efficiency for the social welfare and EVMN, averaged over $10^5$ Rayleigh fading realizations, as a function of $\mathrm{SNR[dB]} = 10 \log_{10} \frac{1}{\sigma^2}$. The figures show the influence of the decoding order on the considered metrics for three choices: increasing order of $|h_i|$ (updated for each packet); decreasing order of $|h_i|$; random decoding order. We see that in contexts where the performance metric is clearly identified ($v$ or $w$), the optimal decoding order can be found and used by the BS, at a price of a certain amount of additional signaling. On the other hand, if there is no dominant arguments in favor of one of them, choosing a random decoding order is relevant. Random decoding order has also two other advantages: 1. The order does not necessarily need to be generated by the BS. For example, in current cellular networks, almost all the mobile phones have an FM receiver. The FM signal can be sampled and thus used a common source of random decoding order. In this case, the additional signaling from the BS is zero; 2. It is in favor of creating fairness. As a second step, we want to compare SIC and SUD. For this, we assume an RCDMA system with $R_i = 100$ kbps for all $i \in \{1, ..., K\}$. The efficiency-function chosen is $f(x) = (1 - e^{-x})^M$ with $M \in \{2, 5, 10, 20, 50, 100\}$. The corresponding values for $\beta^*(M)$ are respectively $1.25, 2.66, 3.61, 4.51, 5.65, 6.47$. Note that $M = 100$ is a typical value of the number of symbols per packet in a cellular system whereas the choice of small values for $M$ is more typical in sensor networks measuring a temperature field (low data rates). Fig. 3 represents the quantity $\frac{w^{\mathrm{SIC}}}{w^{\mathrm{SUD}}} - 1$ in percentage as a function of the spectral efficiency $\alpha = \frac{K}{N}$ for $\mathrm{SNR[dB]} = 6$ and random decoding order. The asymptotes $\alpha_{\max} = \frac{1}{\beta^*(M)} + \frac{1}{N}$ are indicated by (red) dotted lines. The gains are particularly significant when the system load is relatively high i.e., when there is a significant amount of interference to be removed after despreading. In fact, when $\frac{K-1}{N}\beta^* \rightarrow 1^-$ the non-cooperative game becomes dramatically inefficient since the transmit powers at the equilibrium diverge; here, once again we recall that we assume a non-saturated NE at which the users do not exploit all their power. Otherwise, a user who would saturate his power constraint would maximize his utility by transmitting at $P_i^{\max}$. Fig. 4 represents the same type of comparison for the Stackelberg and non-cooperative games. The corresponding results have been obtained by assuming the same scenario just described for Fig. 3 and a random choice of the leader. We observe the same behavior for the relative performance gain, which is in part due to the fact that the non-cooperative game is not designed to operate at a load close to the maximal admissible system load i.e., $\alpha_{\max} = \frac{K-1}{N}\beta^*$. If the load is small or/and the packets are long, the gains brought by the Stackelberg approach are smaller but reasonably high taking into account the additional signaling required. If better gains have to be obtained, the proposed approach can be extended by choosing a group of leaders and a group of followers. In this respect we have shown in Sec. III that, in this case, whatever the channel gain, any follower will always perform better than a follower in terms of energy-efficiency (since any leader of the leading group gets an SINR equal to $\gamma^*$ and any follower of the other group gets an SINR equal to $\beta^*$) and have verified by simulations that there exists an optimal fraction of followers (or leaders) that maximizes social welfare. Note that the gain in terms of individual utilities is easy to deduce from our simulations as we assume a simple scenario where the users have different fading gains but same path losses, the relative gain for each user coincides with the relative gain for the network since the utilities are averaged over the fading gains $\forall i \in \{1, ..., K\}, \mathbb{E}\left[\frac{w^{\mathrm{SIC}}}{w^{\mathrm{SUD}}} - 1\right] = \mathbb{E}\left[\frac{u_i^{\mathrm{SIC}}}{u_i^{\mathrm{SUD}}} - 1\right]$.

## VII. Conclusion

We have analyzed the effect of hierarchy in energy-efficient power control games both on the individual user and overall network performance. The existence and uniqueness of equilibria in the considered games is insured under reasonable assumptions. In fact, when assuming SIC at the receiver the existence and uniqueness of an NE is always guaranteed. We have shown that it is also possible to characterize completely and analytically the efficiency of these equilibria. Compared to most existing analyses conducted in other fields for which game theory is applied, some unusual results have been obtained. In particular it is shown that: both the leader and followers benefit from hierarchy; following is more energy-efficient than leading. Another interesting result is that, when introducing a super-leader (the receiver) in both games considered, the best strategy of the super-leader is strongly related to the choice of the global network efficiency measure. For example, the best decoding order for the social welfare corresponds to the worse decoding order for the EVMN. This shows that implementing a SIC with a random decoding order has two desirable features: if the decoding order is generated from an external source (e.g., an FM signal) there is no additional signaling; choosing the decoding order randomly allows the network to obtain performance gains less dependent on the performance index. Also, after optimization of the social welfare, the super-leader obtains that the users who were "rich" in terms of link quality are now even "richer": this shows that social welfare can be an unfair measure of energy-efficiency of the network. To conclude we would like to mention possible extensions of the presented work: Introduce the concept of classes of leaders and followers to optimize the fraction of followers in the network (e.g., the number of cognitive terminals); Analyze the impact of channel uncertainty on the users' behavior and individual performance; in particular, it would be useful to refine our analysis by considering a non-perfect SIC.

# Appendix A
## Proof of proposition 2

Using the utility function defined by Eq. (3), we obtain from Eq. (4) that for all $p_i$, the optimal decision of a follower $j \neq i$, given the power of the leader, is to choose the power

$$p_j^{\mathrm{SE}}(p_i) = \frac{\beta^*}{1-(K-2)\beta^*}\frac{\sigma^2 + p_i|h_i|^2}{|h_j^2|}. \tag{34}$$

This equation is given by a non-cooperative game among followers where the power of the leader is included in the noise. Note that the $(K-2)\beta^*$ corresponds to the interference generated by the followers $k \in \{1, ..., K\}$ for which $k \neq i$ and $k \neq j$. Plugging $p_j^{\mathrm{SE}}(p_i)$ into the utility expression for user $i$, we obtain:

$$u_i(p_i) = \frac{R_i f\left[\frac{p_i|h_i|^2(1-(K-2)\beta^*)}{\beta^* p_i|h_i|^2(K-1)+\sigma^2(1+\beta^*)}\right]}{p_i} \triangleq \frac{R_i f\left[s(p_i)\right]}{p_i} \tag{35}$$

where we use the function $s(p_i)$ to refer to the SINR of the leader. We have that $p_i^{\mathrm{SE}}$ has to verify $p_i^{\mathrm{SE}} s'(p_i^{\mathrm{SE}}) f'\left[s(p_i^{\mathrm{SE}})\right] = f\left[s(p_i^{\mathrm{SE}})\right]$. This equation is equivalent to finding $p_i$ such that

$$s(p_i)\left[1 - \frac{(K-1)\beta^*}{1-(K-2)\beta^*}s(p_i)\right] f'\left[g(p_i)\right] = f\left[g(p_i)\right], \tag{36}$$

since $p_i g'(p_i) = g(p_i)\left[1 - \frac{(K-1)\beta^*}{1-(K-2)\beta^*}\right]$.

Denote for simplicity by $x$ the quantity $x \triangleq \frac{p_i|h_i|^2(1-(K-2)\beta^*)}{\beta^* p_i|h_i|^2(K-1)+\sigma^2(1+\beta^*)} = s(p_i)$. Studying the existence and uniqueness issues for $p_i$ is equivalent to analyzing those of $x_0$ such that $\phi(x_0) = 0$ with $\phi(x) = x\left[1 - \frac{(K-1)\beta^*}{1-(K-2)\beta^*}x\right] f'(x) - f(x)$ where $f$ have all the properties mentioned in [19] i.e.,

- $f$ is continuous over $[0, +\infty)$ with $f(0) = 0$ and $\lim_{x\to+\infty} f(x) = const$. Recall that $const = 1$ in [8];
- $\forall x \geq 0,\ f'(x) \geq 0$;
- as $f$ is S-shaped we can define an $x_c$ such that $\forall x \leq x_c,\ f''(x) \geq 0$ and $\forall x \geq x_c,\ f''(x) \leq 0$;
- $\lim_{p_i \to 0} u_i(p_i) = 0$.

Therefore our problem boils down to knowing the sign of $\phi'(x)$ for $x \geq 0$.

*Existence of* $x_0$. We know that $\phi(0) = 0$ and $\forall x \geq \frac{(K-1)\beta^*}{1-(K-2)\beta^*},\ \phi(x) < 0$. Therefore if we can prove that $\phi$ is locally strictly positive on the interval $]0, \frac{(K-1)\beta^*}{1-(K-2)\beta^*}[$ the existence of $x_0$ will be guaranteed. A sufficient condition for the existence of $x_0$ is $\frac{f''(0)}{f'(0)} \geq 2\frac{(K-1)\beta^*}{1-(K-2)\beta^*}$. To check this use $\phi''(x) = -2\frac{(K-1)\beta^*}{1-(K-2)\beta^*}f'(x) + f''(x) + x\left[-4\frac{(K-1)\beta^*}{1-(K-2)\beta^*}f''(x) + (1 - \frac{(K-1)\beta^*}{1-(K-2)\beta^*}x)f'''(x)\right]$ and call for the Taylor-Lagrange theorem: there exists $c \in ]0, x[$ such that $\phi(x) = \phi''(0)\frac{x^2}{2} + \phi'''(c)\frac{c^3}{6}$. The quantity $\frac{c^3}{x^2} \leq x$ can be made arbitrary small in the neighborhood of zero. The proposed sufficient condition insures the convexity of $\phi$ and $\phi$ is therefore locally strictly positive.

*Uniqueness of* $x_0$. It follows from the existence and the fact that $\phi$ is assumed to have a single stationary point in the interval $]0;\gamma^*[$.

*Determination of the powers at the SE*. Knowing $x_0 = \gamma^*$ we obtain $p_i^{\mathrm{SE}}$ by using the reciprocal function of $s$, i.e., $p_i^{\mathrm{SE}} = s^{-1}(x_0)$ which gives

$$p_i^{\mathrm{SE}} = \frac{\sigma^2}{|h_i|^2}\frac{\gamma^*(1+\beta^*)}{1-(K-1)\gamma^*\beta^*-(K-2)\beta^*}. \tag{37}$$

From Eq.(34), we obtain the power for follower $j \neq i$:

$$p_j^{\mathrm{SE}} = p_j^{\mathrm{SE}}(p_i^{\mathrm{SE}}) \quad = \quad \frac{\beta^*}{|h_j|^2}\frac{\sigma^2+\sigma^2\frac{\gamma^*(1+\beta^*)}{1-(K-1)\gamma^*\beta^*-(K-2)\beta^*}}{1-(K-1)\gamma^*\beta^*-(K-2)\beta^*} \tag{38}$$

$$= \quad \frac{\sigma^2}{|h_j|^2}\frac{\beta^*(1+\gamma^*)}{1-(K-1)\gamma^*\beta^*-(K-2)\beta^*}. \tag{39}$$

## APPENDIX B
## PROOF OF PROPOSITION 9

Let $p_{tot}^i$ be the total power at the equilibrium when user $i$ is leader. Then for all $j \neq i$,

$$v^{(i)} - v^{(j)} = \frac{(R_j - R_i)\left[f(\beta^*) - f(\gamma^*)\right] + v^{(j)}\frac{\sigma^2(\beta^*-\gamma^*)\left(\frac{1}{|h_i|^2}-\frac{1}{|h_j|^2}\right)}{1-(K-1)\beta^*\gamma^*-(K-2)\beta^*}}{p_{tot}^i}. \tag{40}$$

From the definition of $v$ we have $v^{(i)} = \frac{R_i f(\gamma^*)+\sum_{j\neq i} R_j f(\beta^*)}{p_{tot}^i}$. Thus

$$v^{(i)}p_{tot}^i - v^{(j)}p_{tot}^j = (R_j - R_i)\left[f(\beta^*) - f(\gamma^*)\right]. \tag{41}$$

On the other hand we have that

$$p_{tot}^j - p_{tot}^i = \sum_k p_k^{\mathrm{SE}}(j \equiv \text{leader}) - \sum_k p_k^{\mathrm{SE}}(i \equiv \text{follower}) = \frac{\sigma^2\left(\gamma^*-\beta^*\right)\left(\frac{1}{|h_j|^2}-\frac{1}{|h_i|^2}\right)}{1-(K-1)\beta^*\gamma^*-(K-2)\beta^*}, \tag{42}$$

which can be rewritten as

$$p_{tot}^j = p_{tot}^i + \frac{\sigma^2(\gamma^*-\beta^*)\left(\frac{1}{|h_j|^2}-\frac{1}{|h_i|^2}\right)}{1-(K-1)\beta^*\gamma^*-(K-2)\beta^*}. \tag{43}$$

Plugging the latter expression of $p_{tot}^j$ in (41), we obtain

$$p_{tot}^i(v^{(i)} - v^{(j)}) = (R_j - R_i)\left[f(\beta^*) - f(\gamma^*)\right] + v^{(j)}\frac{\sigma^2(\beta^*-\gamma^*)(\frac{1}{|h_j|^2}-\frac{1}{|h_i|^2})}{1-(K-1)\beta^*\gamma^*-(K-2)\beta^*}. \tag{44}$$

Therefore, we finally have that $v^{(i)} \geq v^{(j)}$ if and only if

$$v^{(j)}\frac{\sigma^2(\beta^*-\gamma^*)\left(\frac{1}{|h_j|^2}-\frac{1}{|h_i|^2}\right)}{1-(K-1)\beta^*\gamma^*-(K-2)\beta^*} \geq (R_i - R_j)\left[f(\beta^*) - f(\gamma^*)\right]. \tag{45}$$

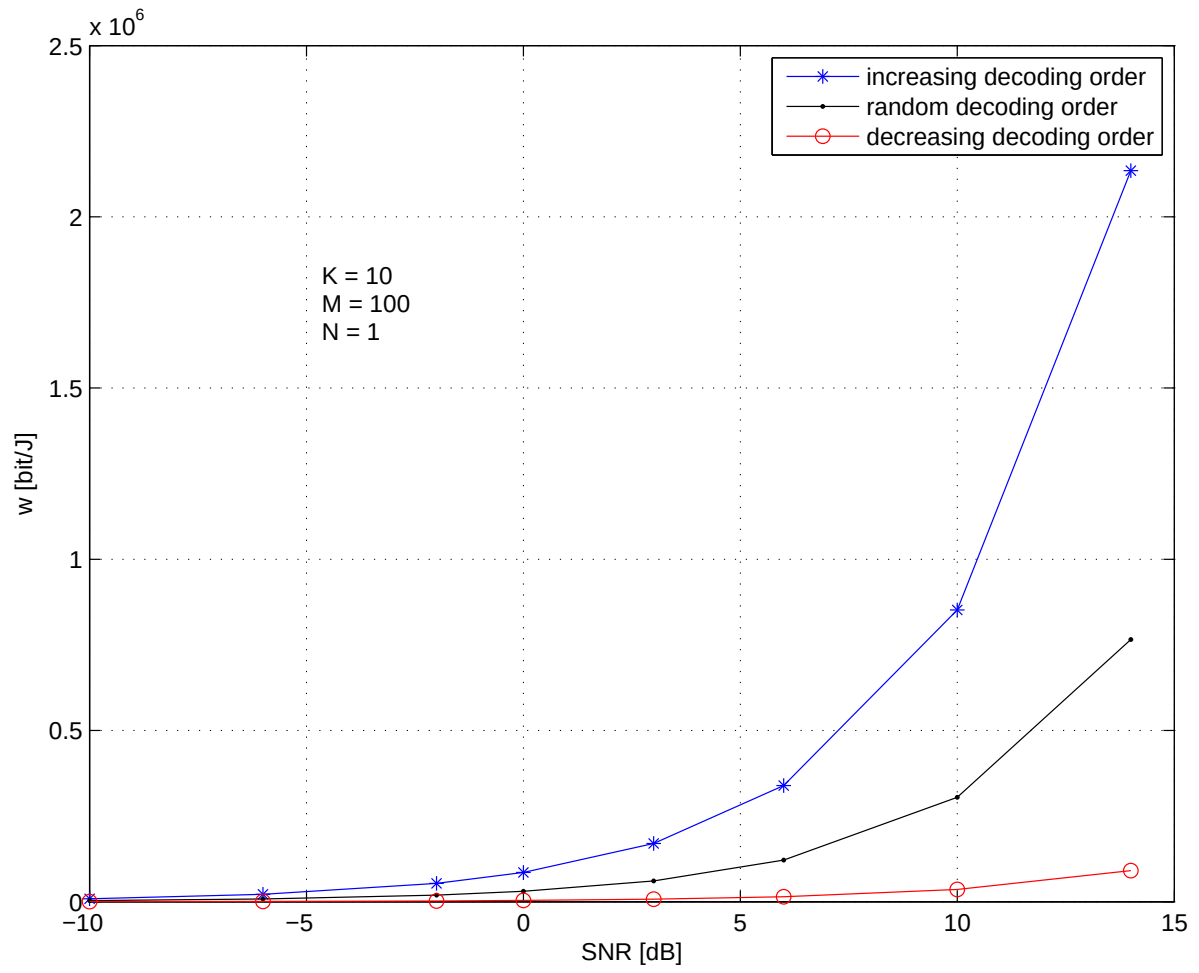

Fig. 1. Influence of the decoding order on the social welfare.

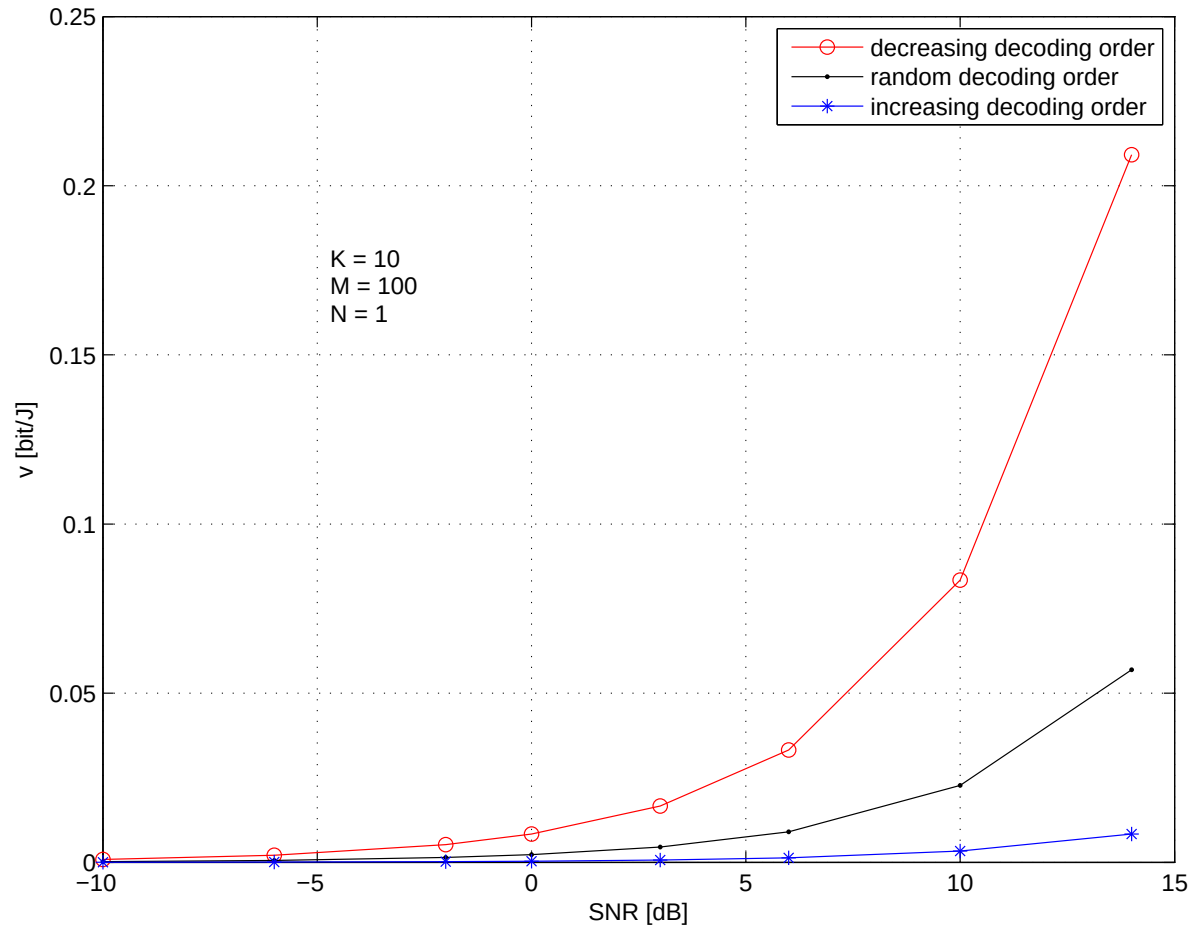

Fig. 2. Influence of the decoding order on the equivalent virtual MIMO network energy efficiency.

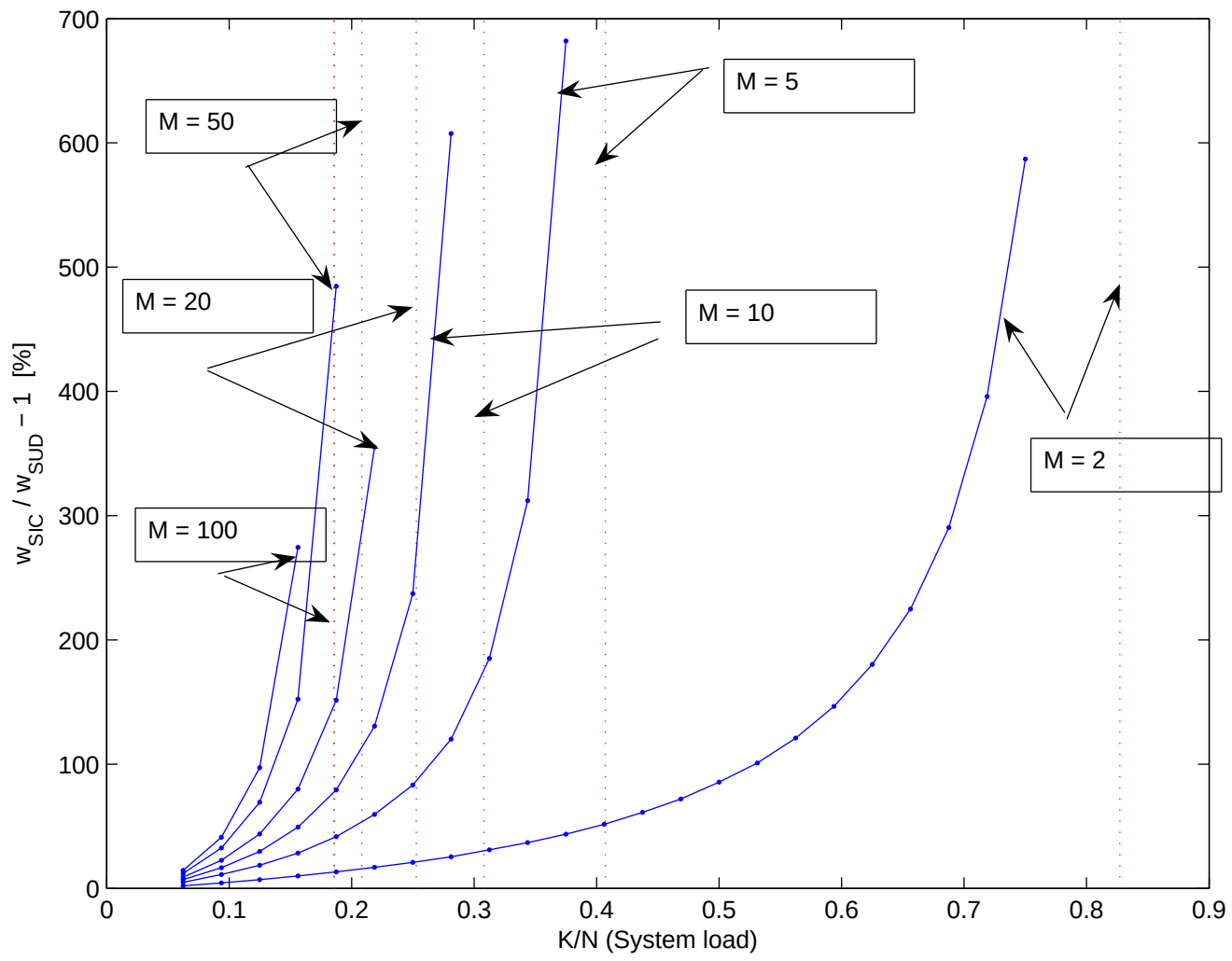

Fig. 3. Network energy-efficiency vs spectral efficiency. Influence of the decoding scheme (SIC/SUD) for different system loads and packet lengths (social welfare); random decoding order.

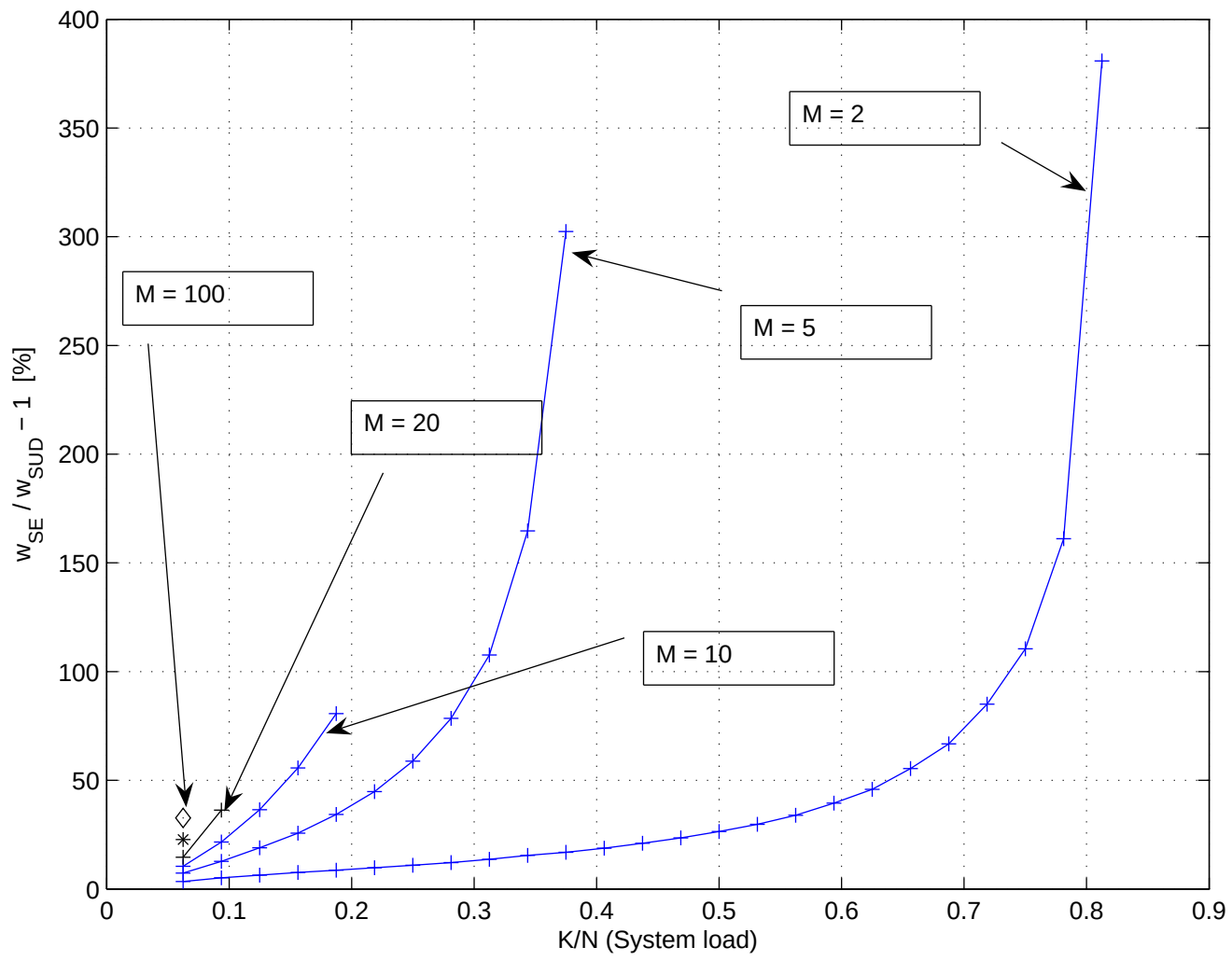

Fig. 4. Network energy-efficiency vs spectral efficiency. Stackelberg game versus non-cooperative game for different system loads and packet lengths (social welfare); random choice of the leader.